\pdfoutput=1
\documentclass[iop]{emulateapj}
\usepackage{amsmath,amssymb,color,verbatim}
\slugcomment{}
\shorttitle{Observational signatures of end-dominated collapse in the S242 filamentary structure}
\shortauthors{L.~K. Dewangan et al.}
\begin{document}
\title{Observational signatures of end-dominated collapse in the S242 filamentary structure}
\author{L.~K. Dewangan\altaffilmark{1}, L.~E. Pirogov\altaffilmark{2}, O.~L. Ryabukhina\altaffilmark{2,3}, D.~K. Ojha\altaffilmark{4}, and I. Zinchenko\altaffilmark{2}}
\email{lokeshd@prl.res.in}
\altaffiltext{1}{Physical Research Laboratory, Navrangpura, Ahmedabad - 380 009, India.}
\altaffiltext{2}{Institute of Applied Physics of the Russian Academy of Sciences, 46 Ulyanov st., Nizhny Novgorod 603950, Russia.}
\altaffiltext{3}{Institute of Astronomy of the Russian Academy of Sciences, 48 Pyatnitskaya st., Moscow 119017, Russia.}
\altaffiltext{4}{Department of Astronomy and Astrophysics, Tata Institute of Fundamental Research, Homi Bhabha Road, Mumbai 400 005, India.}
\begin{abstract}
We present new CO ($^{13}$CO(1--0) and C$^{18}$O(1--0)) and CS(2--1) line observations of an elongated filamentary structure (length $\sim$30 pc) in the star-forming site S242, which were 
taken with the OSO-20m telescope. One filament's end hosts the S242 H\,{\sc ii} region, while the other end contains {\it Planck} cold clumps. 
Several sub-regions are identified in the filament, and are supersonic with Mach number of 2.7--4.0. 
The study of the dynamical states shows supercritical nature of the sub-regions (except central part), 
which could not be supported by a combination of thermal and turbulent motions. 
Young stellar objects are seen toward the entire filament, but more concentrated toward its ends. 
Dense molecular cores are observed mainly toward the filament ends, and are close to virial equilibrium. 
Position-velocity plots trace velocity gradients ($\sim$1 km s$^{-1}$ pc$^{-1}$) toward both the ends. 
An oscillatory pattern in velocity is also observed toward the filament, indicating its fragmentation. 
The collapse timescale of the filament is computed to be $\sim$3.5 Myr. 
Using the $^{13}$CO data, the structure function in velocity of the filament is found to be very 
similar as seen in the Musca cloud for lags $\sim$1--3~pc, and deviates from the Larson's velocity-size relationship.
The observed oscillatory pattern in the structure function at higher lags suggests the existence of large-scale and ordered velocity gradients as well as 
the fragmentation process through accretion along the filament. 
Considering all the observed results along with their uncertainties, the S242 filament is a very good example of the end-dominated collapse.
\end{abstract}
\section{Introduction}
\label{sec:intro}
In the past fifteen years, there has been a remarkable observational progress in the study of star formation with the availability of the infrared and sub-millimeter (sub-mm) data provided by the space-based {\it Spitzer} and {\it Herschel} telescope facilities \citep[e.g.,][]{churchwell06,churchwell07,andre10,andre14}. 
These data sets have unveiled numerous filamentary features, where mid-infrared bubbles/shells associated with 
H\,{\sc ii} regions, embedded clumps, and clusters of young stellar objects (YSOs) are commonly identified \citep[e.g.,][]{deharveng10,andre14}. 
These observational evidences have encouraged investigators to study the physical mechanisms of filament fragmentation, and the role of filaments in the formation of dense massive star-forming clumps and young stellar clusters \citep[e.g.,][]{myers09,andre10,deharveng10,schneider12,baug15,dewangan15,dewangan16,dewangan16b,dewangan17a,
dewangan17b,dewangan17c,dewangan17d,dewangan17e,contreras16}, which are still debated.  
In this context, some existing models for the long, but finite-sized, filaments predict that the fragmentation and collapse can take place at the ends of the filaments (i.e., end-dominated collapse), where the gas has an enhanced acceleration \citep[e.g.,][]{bastien83,burkert04,heitsch08,pon11,pon12,clarke15}. However, observational assessment of such existing numerical simulations is very limited in the literature \citep[e.g.,][]{zernickel13,beuther15,kainulainen16,hacar16}. It requires a promising sample of the filamentary features and the knowledge of molecular gas motion in the direction of these objects \citep[e.g.,][]{kainulainen16,contreras16,williams18}. However, high-resolution molecular line observations of several interesting filamentary features revealed by the space-based telescopes are not available. In this connection, an embedded filamentary structure in a massive star-forming region, LBN 182.30+00.07 or Sh 2-242 (hereafter S242) can be considered as one of such sources \citep[e.g.,][hereafter Paper~I]{dewangan17c}, and is the target source of the present study.  
To examine the dynamics and physical conditions of the gas in the S242 filament, an extensive analysis of new molecular (CO and CS) line observations taken with the 20-m telescope of the Onsala Space Observatory (OSO) is carried out in this paper. Such investigation has also allowed us to examine the existing numerical predictions concerning the fragmentation of filamentary feature. 

Following the introduction in this section, we give an overview of the site S242 
in Section~\ref{sec:intro2}. In Section~\ref{sec:obser}, we present the molecular line observations and data reduction.
Observational outcomes are presented in Section~\ref{sec:data}, and are discussed 
in Section~\ref{sec:disc}. The main conclusions of this paper are summarized in Section~\ref{sec:conc}.
\section{Overview of the site S242}
\label{sec:intro2}
\begin{figure*}
\epsscale{1.1}
\plotone{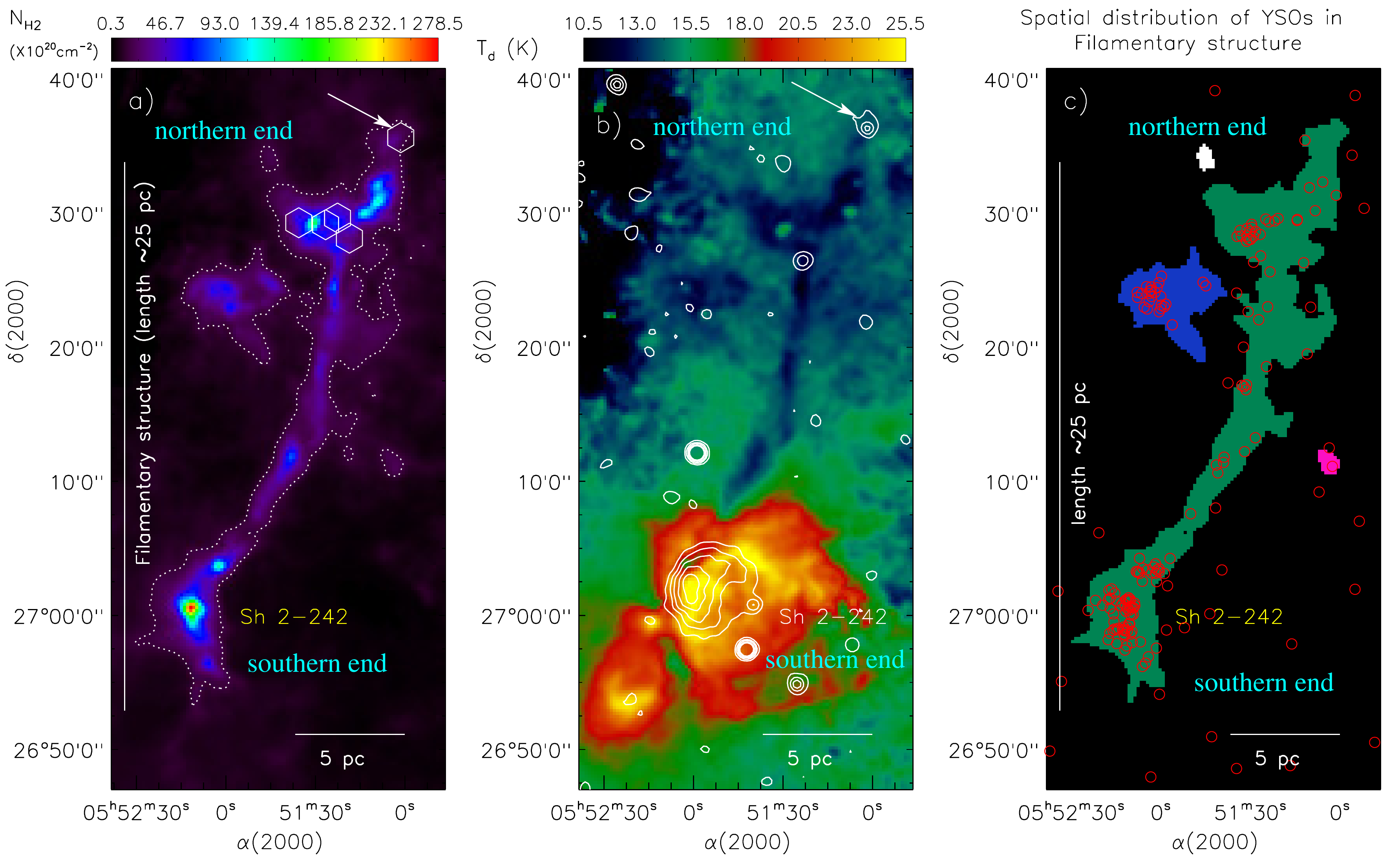}
\caption{Physical environment of the site S242. 
a) {\it Herschel} column density ($N(\mathrm H_2)$) map of the region around S242. 
A filamentary structure is traced in the column density map at a contour level of 1.5 $\times$ 10$^{21}$ cm$^{-2}$ (see dotted contours). 
The positions of five PGCCs \citep[e.g.,][]{yuan16} are also highlighted by hexagons. 
An arrow indicates the location of the PGCC G181.84+00.31a1 \citep[e.g.,][]{yuan16}. 
b) Overlay of the NVSS 1.4 GHz radio continuum emission contours (in white) on the {\it Herschel} temperature map. 
The NVSS 1.4 GHz continuum contours are shown with the levels of 0.45 mJy/beam $\times$ [3, 11, 20, 33, 44]. 
An arrow indicates the location of the radio continuum source NVSS 055101+273627 \citep[e.g.,][]{condon98}. 
c) Spatial distribution of YSOs (see red circles) in the direction of filament as seen in the {\it Herschel} column density map (see Figure~\ref{fg1}a). All these results are taken from Paper~I. The scale bar corresponding to 5 pc (at a distance of 2.1 kpc) is shown in each panel.}
\label{fg1}
\end{figure*}

Situated at a distance of 2.1$\pm$0.7~kpc \citep{blitz82}, the site S242 is associated with a filamentary structure, 
which has been revealed by the {\it Herschel} sub-mm data (e.g., Paper~I). 
Based on the published results from Paper~I, Figures~\ref{fg1}a and~\ref{fg1}b display the 
{\it Herschel} column density and temperature maps (resolution $\sim$37$''$) of the site S242, respectively. 
A variation in the dust temperature is evident along the major axis of the filament (see Figure~\ref{fg1}b). 
The locations of the ionized emission traced in the NVSS 1.4 GHz continuum map are found in Figure~\ref{fg1}b. 
The southern end of the filament has been found to host the S242 H\,{\sc ii} region, where an extended structure in the {\it Herschel} temperature map is depicted with a temperature (T$_{d}$) range of $\sim$19--26 K (see Figure~\ref{fg1}b). 
The S242 H\,{\sc ii} region was found to be powered by a star BD+26 980 of spectral type B0.5V--B0V \citep{hunter90}, and has a dynamical age of $\sim$0.5 Myr (see also Paper~I). In Paper~I, the photometric 1--5 $\mu$m data of point-like sources were analyzed to identify YSOs in the site S242, which are also shown in Figure~\ref{fg1}c. 
Both the northern and southern ends of the S242 filament have been found with the noticeable star formation activities compared to its other parts (see Figure~\ref{fg1}c). In the southern direction, a molecular CO outflow toward IRAS 05490+2658, which is situated $\sim$5$\arcmin$ east of the S242 H\,{\sc ii} region, was reported by \citet{snell90} \citep[see also][]{varricatt10}.  
Several {\it Herschel} clumps/condensations have been identified along the filament (see Figure~\ref{fg1}a). The most massive clumps (M$_{clump}$ $\sim$250--1020 M$_{\odot}$) have also been detected toward both the filament ends (see Table~1 in Paper~I). 
It was also found that the total mass of three {\it Herschel} clumps (i.e. $\sim$1700 M$_{\odot}$) 
seen in the southern filament end was almost equal to the total mass of three clumps (i.e. $\sim$1750 M$_{\odot}$) observed 
in the northern filament end (see Table~I in Paper~I). 
Based on the NANTEN $^{13}$CO(1--0) low-resolution map (beam size $\sim$2$'$.7), 
molecular gas toward the site S242, having a radial velocity (V$_{lsr}$) of $\sim$0.7 km s$^{-1}$, was studied by \citet{kawamura98}. To the best of our knowledge, there are no high-resolution molecular line observations reported toward the S242 filament. In Paper~I, the observed results in the site S242 were explained by the end-dominated collapse process. However, the study related to the internal dynamical properties/internal kinematics of the emitting gas in the filament is yet to be performed, which is essential to further explore the ongoing physical process in the site S242. 
\section{Data and analysis}
\label{sec:obser}
\subsection{New Observations}
New molecular line observations were carried out in March 2018 with the 20-m telescope of the OSO (project O2017a-03). 
We used the 3-mm dual-polarization receiver with a SIS mixer, which has a noise temperature of $\sim$45-60 K in the 85-116~GHz frequency range \citep{belitsky15}. 
Fast Fourier Transform spectrum analyser with a 2.5~GHz bandwidth and a 76~kHz frequency resolution (32768 channels) was used in the observational experiment. 
The CO ($^{13}$CO(1--0) and C$^{18}$O(1--0)) and the CS(2--1) lines were simultaneously observed in the upper (i.e. 110~GHz) 
and lower (i.e. 98~GHz) sidebands, respectively.
The spectrometer gives a velocity resolution of $\sim$0.21-0.23~km s$^{-1}$ at these frequencies. 
To obtain high sensitivity data, the signals from two polarizations were added during the reduction process, which provided an increase of sensitivity by a factor of $\sqrt{2}$.
Observations were performed in the frequency switching mode. 
The system noise temperature during most of the observations varied in ranges of $\sim$160-300 K (at 110~GHz) 
and $\sim$115-200 K (at 98~GHz) depending on the elevation of the target.
Main beam efficiency depends on the source elevation, and varies in the ranges of $\sim$0.32-0.45 (at 110~GHz) 
and 0.38-0.50 (at 98~GHz) for elevations of $\sim$25$\degr$--60$\degr$. These parameters were used to convert antenna temperatures to main beam brightness temperatures. The half power beam widths of the OSO-20m antenna are $\sim$35$''$  (at 110~GHz) 
and $\sim$39$''$ (at 98 GHz). Pointing and focus were regularly checked by the observations of SiO masers.
Mapping was done with 20$''$ grid spacing. 
The line data were processed using the CLASS program from the GILDAS\footnote[1]{http://www.iram.fr/IRAMFR/GILDAS} package, 
the XS software\footnote[2]{https://www.chalmers.se/en/researchinfrastructure/oso/radio-astronomy/Pages/software.aspx}, and our original programs.
\subsection{Others}
The archival data sets in the radio and sub-mm regimes were obtained from publicly available surveys (e.g., the NRAO VLA Sky Survey \citep[NVSS; $\lambda$ = 21 cm; resolution = 45$\arcsec$;][]{condon98}, {\it Planck\footnote[3]{{\it Planck} (http://www.esa.int/Planck) is a project of the 
European Space Agency (ESA) with instruments provided by two scientific consortia funded by ESA member states and led by 
Principal Investigators from France and Italy, telescope reflectors provided through a collaboration between ESA and a 
scientific consortium led and funded by Denmark, and additional contributions from NASA (USA).} intensity 
image \citep[$\lambda$ = 850 $\mu$m; resolution = 300$''$;][]{planck14} observed with the High Frequency 
Instrument (HFI) at 353 GHz,}  and the {\it Herschel} Infrared Galactic Plane Survey \citep[Hi-GAL; $\lambda$ = 70, 160, 250, 350, 500 $\mu$m; resolutions = 5$''$.8, 12$''$, 18$''$, 25$''$, 37$''$;][]{molinari10}). 
We also used the published results from Paper~I.
\section{Results}
\label{sec:data}
\subsection{S242 site: {\it Planck} Galactic Cold Clumps}
\label{ssec:data1}
In the direction of the northern filament end, at least five {\it Planck} Galactic Cold Clumps \citep[PGCCs; from][]{yuan16} are reported (see Figure~\ref{fg1}a). It is thought that PGCCs may represent the early stages of star formation \citep[e.g.,][]{tang18}. 
Using the HCN (J = 1--0) and HCO$^{+}$ (J = 1--0) line observations, \citet{yuan16} 
studied dense gas toward the PGCCs in a velocity (V$_{lsr}$) range of 1.95--2.2 km s$^{-1}$. The positions of PGCCs G181.84+00.31a1, 
G181.84+00.31a2, G181.84+00.31b1, G181.84+00.31b2, and G182.04+00.41b1 are shown in Figure~\ref{fg1}a \citep[see][for more details]{yuan16}. 

We find a radio continuum source NVSS 055101+273627 \citep[e.g.,][]{condon98} 
toward the PGCC G181.84+00.31a1 (see an arrow in Figures~\ref{fg1}a and~\ref{fg1}b), while other remaining PGCCs are not associated with the NVSS radio continuum 
emission (1$\sigma$ $\sim$0.45 mJy beam$^{-1}$).  
Using the NVSS radio continuum data, the source NVSS 055101+273627 is found to be powered by a B1V--B0.5V type star, and its 
dynamical age is estimated to be $\sim$0.1 Myr. 
In Figure~\ref{fg1}c, a majority of the YSOs are found toward both the filament ends (i.e., the S242 H\,{\sc ii} region and PGCCs). 
The clumps traced in the southern filament end containing the S242 H\,{\sc ii} region are found with relatively the warm dust emission compared to the clumps observed in the northern 
filament end hosting the PGCCs (see Figure~\ref{fg1}b). It seems that the clumps distributed toward both the filament ends are probably at different evolutionary stages.
\subsection{S242 site: Spatial distribution of molecular gas}
\label{ssec:data2x}
\begin{figure*}
\epsscale{0.92}
\plotone{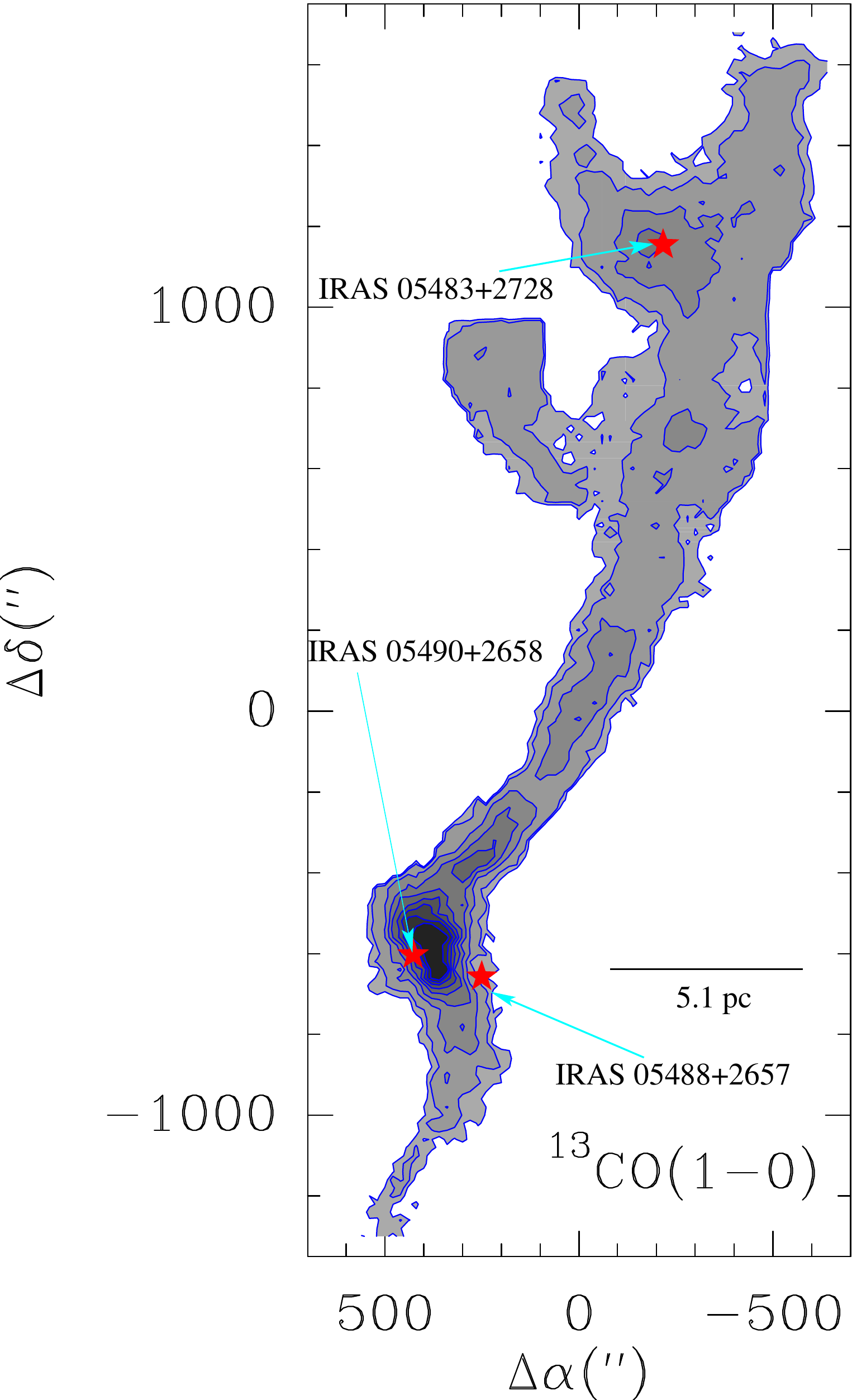}
\caption{Contour map of the $^{13}$CO(1--0) integrated intensity emission in the direction of the site S242.
The contours levels (in K~km~s$^{-1}$) range from 5 to 40 with a step of 5 with an additional contour of 3 K~km~s$^{-1}$. The molecular emission is integrated 
over a velocity range of [$-$12, 6] km s$^{-1}$. The axes are offsets with respect to a central position (i.e., RA (2000) = 5h 52m 12.9s; 
Dec (2000) = 26$\degr$ 59$'$ 33$''$). The positions of IRAS 05490+2658, IRAS 05488+2657, and IRAS 05483+2728 
are denoted by stars (in red).}
\label{fg2}
\end{figure*}

Figure~\ref{fg2} shows an intensity map of $^{13}$CO(1--0) integrated over a velocity interval of [$-$12, 6] km s$^{-1}$. 
Interestingly, the elongated appearance of filamentary feature in the site S242, having length $\sim$30 pc, is traced in the molecular map as seen in the {\it Herschel} column density 
map (see Figure~\ref{fg1}a). Several molecular condensations in the filament can also be identified in Figure~\ref{fg2}. 
The IRAS sources located toward both the filament ends are also shown in Figure~\ref{fg2}.

\begin{figure*}
\epsscale{1}
\plotone{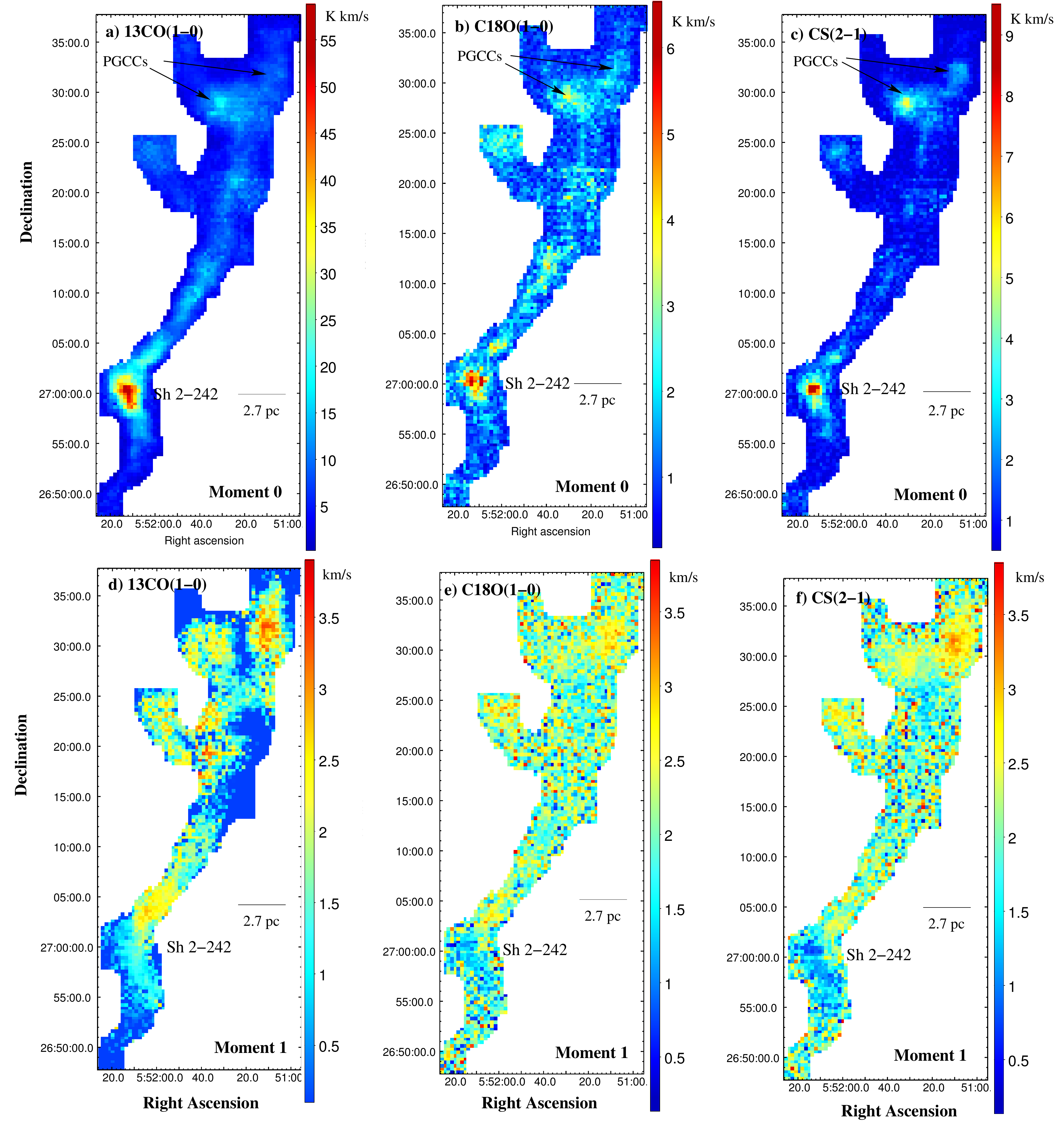}
\caption{a) The $^{13}$CO(1--0) integrated intensity map in the direction of the site S242 (i.e., Moment-0 map). b) Moment-0 map of C$^{18}$O(1--0). 
c) Moment-0 map of CS(2--1). d) Moment-1 map of $^{13}$CO. e) Moment-1 map of C$^{18}$O. d) Moment-1 map of CS(2--1). 
In all the moment-0 maps, the vertical bar shows the color-coded intensity in K km s$^{-1}$. In each moment-1 map, the vertical bar shows the color-coded velocity in km s$^{-1}$.}
\label{fg3}
\end{figure*}

We have also presented integrated intensity (moment-0) maps of $^{13}$CO(1--0), C$^{18}$O(1--0), and CS(2--1) emission 
in Figures~\ref{fg3}a,~\ref{fg3}b, and~\ref{fg3}c, respectively. 
The $^{13}$CO moment-0 map is shown here only for comparison (see also Figure~\ref{fg2}). 
The C$^{18}$O line is a tracer of total molecular gas column density, 
while the CS line is known to depict denser molecular gas compared to the $^{13}$CO data. 
Hence, the denser parts in the filament can be spatially examined in Figures~\ref{fg3}b and~\ref{fg3}c. 
The dense gas traced in the C$^{18}$O(1--0) and CS(2--1) maps is found toward both the filament ends (i.e. the S242 H\,{\sc ii} region and all the PGCCs). 
Figures~\ref{fg3}d,~\ref{fg3}e, and~\ref{fg3}f show moment-1 (velocity field) maps of the $^{13}$CO(1--0), C$^{18}$O(1--0), 
and CS(2--1), respectively. 
With the help of moment-1 map, some signatures of velocity gradients are seen at both the ends of 
the filament (see Figure~\ref{fg3}d). 

\begin{figure*}
\epsscale{1}
\plotone{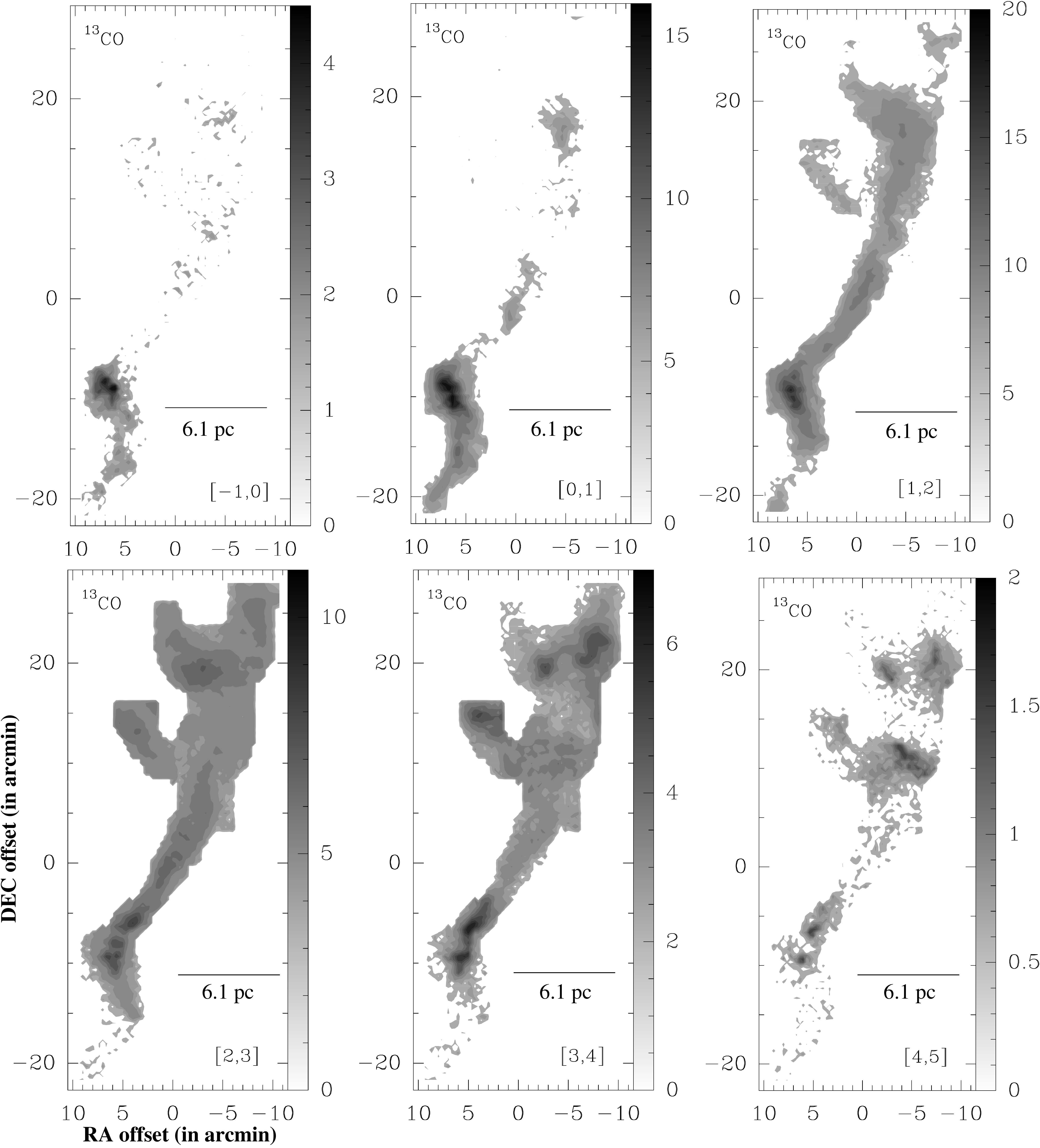}
\caption{The $^{13}$CO(1--0) integrated velocity channel maps (starting from $-$1 km s$^{-1}$ 
at intervals of 1 km s$^{-1}$) in the direction of the site S242. In each panel, the vertical bar shows the color-coded intensity in K km s$^{-1}$.} 
\label{fg4}
\end{figure*}

\begin{figure*}
\epsscale{1}
\plotone{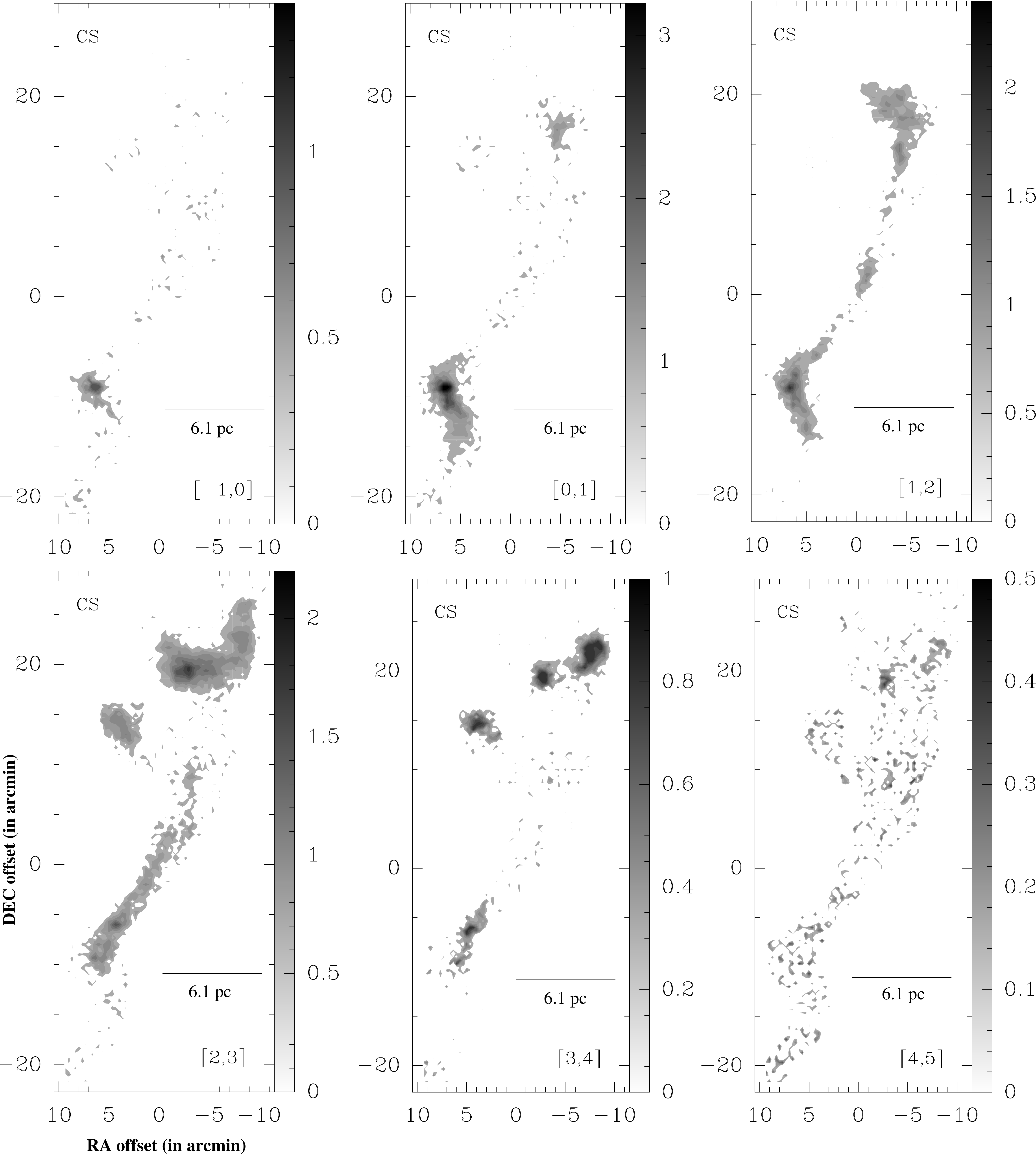}
\caption{The CS(2--1) integrated velocity channel maps (starting from $-$1 km s$^{-1}$ 
at intervals of 1 km s$^{-1}$) in the direction of the site S242. In each panel, the vertical bar shows the color-coded intensity in K km s$^{-1}$.} 
\label{fg5}
\end{figure*}

Figures~\ref{fg4} and~\ref{fg5} display the $^{13}$CO and CS(2--1) velocity channel maps from $-$1 to 5 km s$^{-1}$, respectively. 
Each channel map is obtained by integrating the emission over 1 km s$^{-1}$ velocity intervals. 
These channel maps help us to examine the gas distribution in the filament for a given velocity interval. 
In this paper, the molecular gas in the filament is mainly studied in the velocity interval of [$-$1, 5] km s$^{-1}$. 
However, we find an additional velocity component at [$-$12, $-$6] km s$^{-1}$ toward the northern part of the filament  
using the $^{13}$CO line data (not shown in this paper). In this velocity range, there is no C$^{18}$O(1--0) and CS(2--1) emission detected toward the filament. 
It implies that the gas at [$-$12, $-$6] km s$^{-1}$ has low density. 
In the position-velocity (pv) plot, we find that the velocity component at [$-$12, $-$6] km s$^{-1}$ is kinematically unrelated 
to the one at the [$-$1, 5] km s$^{-1}$ (not shown in this paper). Furthermore, the emission at [$-$12, $-$6] km s$^{-1}$ appears 
to be more extended to the north of the filament, which is not covered in this observational work. 
In order to make any further conclusion on the kinematical structure at [$-$12, $-$6] km s$^{-1}$, one needs to 
obtain further observations for a wide-scale area containing the north part of the filament. 

In Figure~\ref{fg8}a, an arbitrarily chosen solid curve is highlighted on the integrated intensity map of $^{13}$CO emission, 
where the pv, V$_{lsr}$, and $\Delta V$ plots are calculated (see Figures~\ref{fg9}a,~\ref{fg9}b, and~\ref{fg10}). 
The curve indicates the major axis of the filament, and is seen in the direction of the S242 H\,{\sc ii} region, 
the northern cluster of YSOs, and the positions of IRAS 05483+2728 and IRAS 05490+2658 (see Figures~\ref{fg1}c and~\ref{fg2}). 
\subsubsection{Sub-regions in S242 site}
\label{ssec:data3x}
\begin{figure*}
\epsscale{1.15}
\plotone{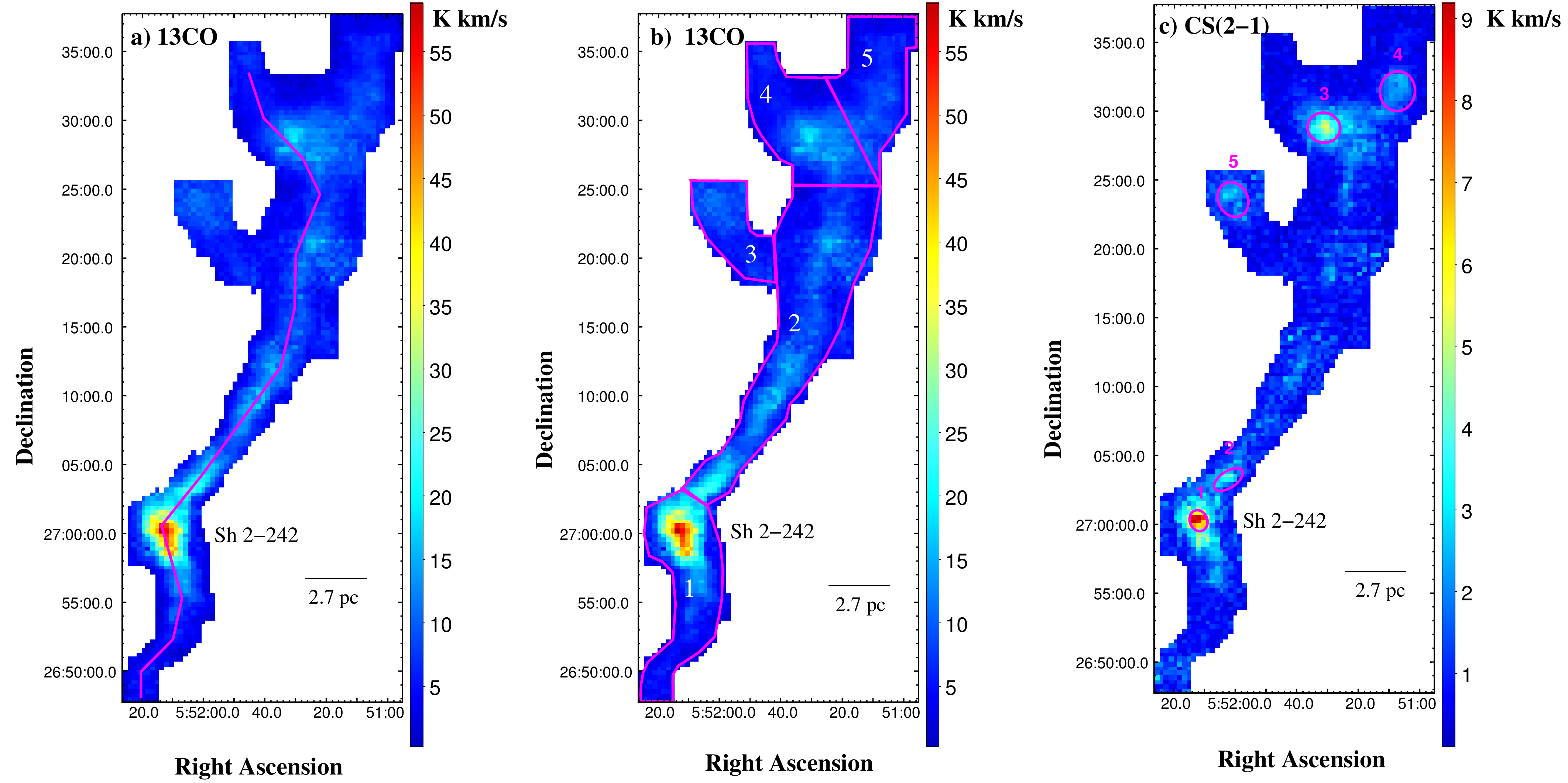}
\caption{a) Overlay of an arbitrarily chosen solid curve/major axis (in pink) on the map of $^{13}$CO(1--0) integrated intensity, where position-velocity plots are calculated (see Figures~\ref{fg9}a, \ref{fg9}b, and~\ref{fg10}). 
b) Overlay of different sub-regions on the map of $^{13}$CO(1--0) integrated intensity, where masses are estimated (see Table~\ref{tab1}). c) Five dense cores highlighted in the map of CS(2--1) integrated intensity (see ellipses). In each panel, the vertical bar at the right shows the color-coded intensity in K km s$^{-1}$.} 
\label{fg8}
\end{figure*}

According to the visual analysis of the $^{13}$CO intensity map, 
several sub-regions are outlined for computing the molecular gas masses (see Figure~\ref{fg8}b). 
We have used the optically thin C$^{18}$O line data to compute the mass of each selected sub-region in the S242 site. Masses ($M_{\rm subreg}$) are obtained by integrating the column density $N(\mathrm H_2)$ over 
the outlined sub-regions (see Figure~\ref{fg8}b) and multiplying by 2.8 $\times$ m$_H$, where m$_H$ is the hydrogen atom mass. 
To obtain $N(\mathrm H_2)$, we have calculated $N(\mathrm C^{18}O)$ from the C$^{18}$O integrated intensities in the optically thin and the local thermodynamic equilibrium (LTE) approximations using an expression from \citet{mangum16}, which is given by
\begin{multline}
N(C^{18}O) = \frac{2.48\times 10^{14} \left(T_{ex}+0.88\right)
\exp\left(\frac{5.27}{T_{ex}}\right)}{\exp\left(\frac{5.27}{T_{ex}}\right)
  - 1} \\
\times\left[\frac{\int T_R\ dv(km\ s^{-1})}{f\left(J_\nu(T_{ex}) - J_\nu(T_{bg})\right)}\right]~\textrm{cm}^{-2}
\label{eq:ntotc18ot}
\end{multline}
where T$_{ex}$ is the excitation temperature which is adopted to be 10~K, $f$ (= 1) is the main beam filling factor 
(i.e., the fraction of the spatial resolution of the measurement filled by the source), T$_{bg}$ (= 2.73~K) is the cosmic background temperature, 
$J(T)=\frac{h\nu}{k}/(\exp(\frac{h\nu}{kT})-1)$, $h$ is the Planck constant, $\int\,T_R\,dv$ is the integral over 
the line profile. Considering a column density ratio [C$^{18}$O]/[H$_{2}$] equal to 1.7 $\times$ 10$^{-7}$ \citep{frerking82}, 
the values of $N(\mathrm H_2)$ are calculated. 
The values of $M_{\rm subreg}$ and length of sub-regions are tabulated in Table~\ref{tab1}. 
These values also allow us to estimate the observed line mass ($M_{\rm line,obs}$) of each sub-region (see Table~\ref{tab1}). 

We have also obtained the critical line mass ($M_{\rm line,crit}$) of a filament. 
For the case of an infinite, isothermal cylinder in equilibrium between thermal and gravitational pressures, the expression of $M_{\rm line,crit}$ \citep{ostriker64} is equal to $2 c^{2}_{\rm s}/G$ $\sim$16~M$_{\odot}$ pc$^{-1}$ $\times$ (T/10 K); where $c_{\rm s}$ is the isothermal sound speed at T(K) and $G$ is the gravitational constant. The contribution of the non-thermal microturbulent gas motions can be included in the calculation of the
critical line mass, which is also known as the virial line mass \citep[i.e.,][]{andre14,kainulainen16}. 
In order to obtain the expression of the virial line mass \citep[i.e. $M_{\rm line,vir}$ = $2 c^{2}_{\rm s, eff}/G$;][]{andre14,kainulainen16}, the sound speed is replaced by an effective sound speed ($c_{\rm s, eff}$ = ($c^{2}_{\rm s}$ + $\sigma^{2}_{\rm NT}$)$^{1/2}$; where $\sigma_{\rm NT}$ is the non-thermal velocity dispersion) in the equation of the critical line mass. We can further write the expression of the virial line mass as: 
\begin{equation}
M_{\rm line,vir} = \left[1 + \left(\frac{\sigma_{\rm NT}}{c_{s}}\right)^2 \right] \times \left[16~M_{\odot}~pc^{-1} \times \left(\frac{T}{10~K}\right) \right]
\label{shh1}
\end{equation}
The non-thermal velocity dispersion is defined by:
\begin{equation}
\sigma_{\rm NT} = \sqrt{\frac{\Delta V^2}{8\ln 2}-\frac{k T_{kin}}{30 m_H}} = \sqrt{\frac{\Delta V^2}{8\ln 2}-\sigma_{\rm T}^{2}} ,
\label{sigmanonthermal}
\end{equation}
where $\Delta V$ is the measured Full Width Half Maximum (FWHM) linewidth of the observed C$^{18}$O spectra, T$_{kin}$ is the gas kinetic temperature (i.e. 10 K), and $\sigma_{\rm T}$ (= $(k T_{kin}/30 m_H)^{1/2}$) is the thermal broadening for C$^{18}$O at T$_{kin}$. 
Mach number (i.e. ratio of $\sigma_{\rm NT}$/$c_{s}$) is also estimated for each sub-region, 
where $c_{s}$ is defined earlier ($c_{s}$ = $(k T_{kin}/\mu m_{H})^{1/2}$ = 0.19 km s$^{-1}$ for T$_{kin}$ = 10 K 
and mean molecular weight ($\mu$) =2.33). The observed Mach number range is computed to be 2.7--4.0, indicating that all the sub-regions are supersonic. Table~\ref{tab1} also lists the linewidth ($\Delta V$ (C$^{18}$O)), $\sigma_{\rm NT}$, and $M_{\rm line,vir}$ of each sub-region. 

The uncertainties of the observed line masses mainly depend on the uncertainty of distance to 
S242 \citep[$\sim$30\%;][]{blitz82}. The adopted value of the excitation temperature (T$_{ex}$ = 10~K) could also affect the values of N(C$^{18}$O).
If we adopt the lower T$_{ex}$ (= 5~K) or higher T$_{ex}$ (= 20~K) in the calculation 
then the line mass is computed to be about 1.2--1.4 times the one at T$_{ex}$ = 10~K. 
Here, one may also keep in mind that the [C$^{18}$O]/[H$_2$] ratio for S242 could be about 1.5 times low 
due to the [$^{16}$O]/[$^{18}$O] dependence on galactocentric distance \citep[e.g.,][]{wilson94}. 
Hence, due to these factors, the observed line mass estimates could be underestimated twice.

The uncertainties of virial line masses depend on the value of mean gas kinetic temperature of sub-regions ($\sim$10~K) and 
the uncertainties of C$^{18}$O FWHM. 
The uncertainties of C$^{18}$O FWHM calculated from the Gaussian fits lead to the uncertainties of $\sim$4--14\% in 
$M_{\rm line,vir}$. 
In the direction of the S242 H\,{\sc ii} region/sub-region~1, the dust temperature (T$_{d}$) range is found to be $\sim$19--26 K (see Figure~\ref{fg1}b). 
If we choose T$_{kin}$ = 20 K in the calculation of $M_{\rm line,vir}$ then it leads the uncertainty of $\sim$10\% in 
$M_{\rm line,vir}$. 
Note that the observed line masses and virial line masses are calculated with an assumption of a zero inclination of the filament, 
which is unknown. Hence, these both observed values represent upper limits. 
Taking into account these considerations, it is possible to state that the observed line masses (mass-to-length ratios) of different parts of the S242 filament are systematically higher than $M_{\rm line,vir}$, implying that these sub-regions (except sub-region ``2" or central part of the filament) could be unstable (see Table~\ref{tab1}).

Furthermore, using the CS(2-1) and C$^{18}$O maps, at least five dense cores are selected in the filament.
These cores are indicated by ellipses on the CS(2--1) map (Figure~\ref{fg8}c). 
The deconvolved mean size of each ellipse is also computed by the Gaussian fitting \citep[see][]{pir03}. 
Two compact cores (sizes $\sim$0.8--0.9 pc) are located near the H\,{\sc ii} region in the direction of the southern filament end. 
A dense core ``3" (size $\sim$1.3 pc; see Figure~\ref{fg8}c) in the northern part of the filament seems 
to be associated with PGCC G182.04+0042b1. The observed cores are also associated with the {\it Herschel} clumps of higher sizes (see Paper~I).
A dense core (see ID ``5" in Figure~\ref{fg8}c) is also embedded in the sub-region ``3". 
The masses of the cores derived from the 
C$^{18}$O(1-0) integrated intensities are $\sim$250--350 M$_{\odot}$. 
We have also computed the virial mass \citep[M$_{vir}$ ($M_\odot$)\,=\,k\,D$_{c}$\,$\Delta V^2$;][where the parameter k\ =\ 105 for 
spherically-symmetric core/clump/cloud with constant density, no external pressure, and no magnetic fields]{maclaren88} 
of diameter D$_{c}$ (in pc) and linewidth $\Delta V$ (in km s$^{-1}$). 
Table~\ref{tab2} contains the derived physical parameters of the cores (i.e., position, diameter (D$_{c}$), mass ($M_{\rm c}$), 
linewidth (CS $\Delta V$), M$_{vir}$, and mean volume density ($\bar n$). 
The cores ``1" and ``2" located near the S242 H\,{\sc ii} region have mean densities several times higher than that of the northern cores ``3--5". The uncertainties of the core masses also depend on the uncertainties of distance to S242, [C$^{18}$O]/[H$_2$] ratio, 
CS FWHM, sizes derived from the Gaussian fits, and excitation temperature. 
Considering all these factors, the uncertainties of the virial masses of the cores lie in the range of $\sim$30--60\%. 
Taking into account the uncertainties of virial mass calculations, the core mass estimates are considered to be close to virial ones, which are typical for star-forming cores. 

Note that the sub-region ``3" contains a dense core and noticeable YSOs, revealing an ongoing star formation in this sub-region. 
However, the sub-region ``3" appears away from the major axis of the filamentary structure (see Figures~\ref{fg1}a and~\ref{fg8}b). 
Hence, we have not further discussed the results of this particular sub-region in this paper.
\subsection{Position-velocity plots}
\label{ssec:data3}
\subsubsection{Velocity field}
\label{xssec:data3}
\begin{figure*}
\epsscale{1}
\plotone{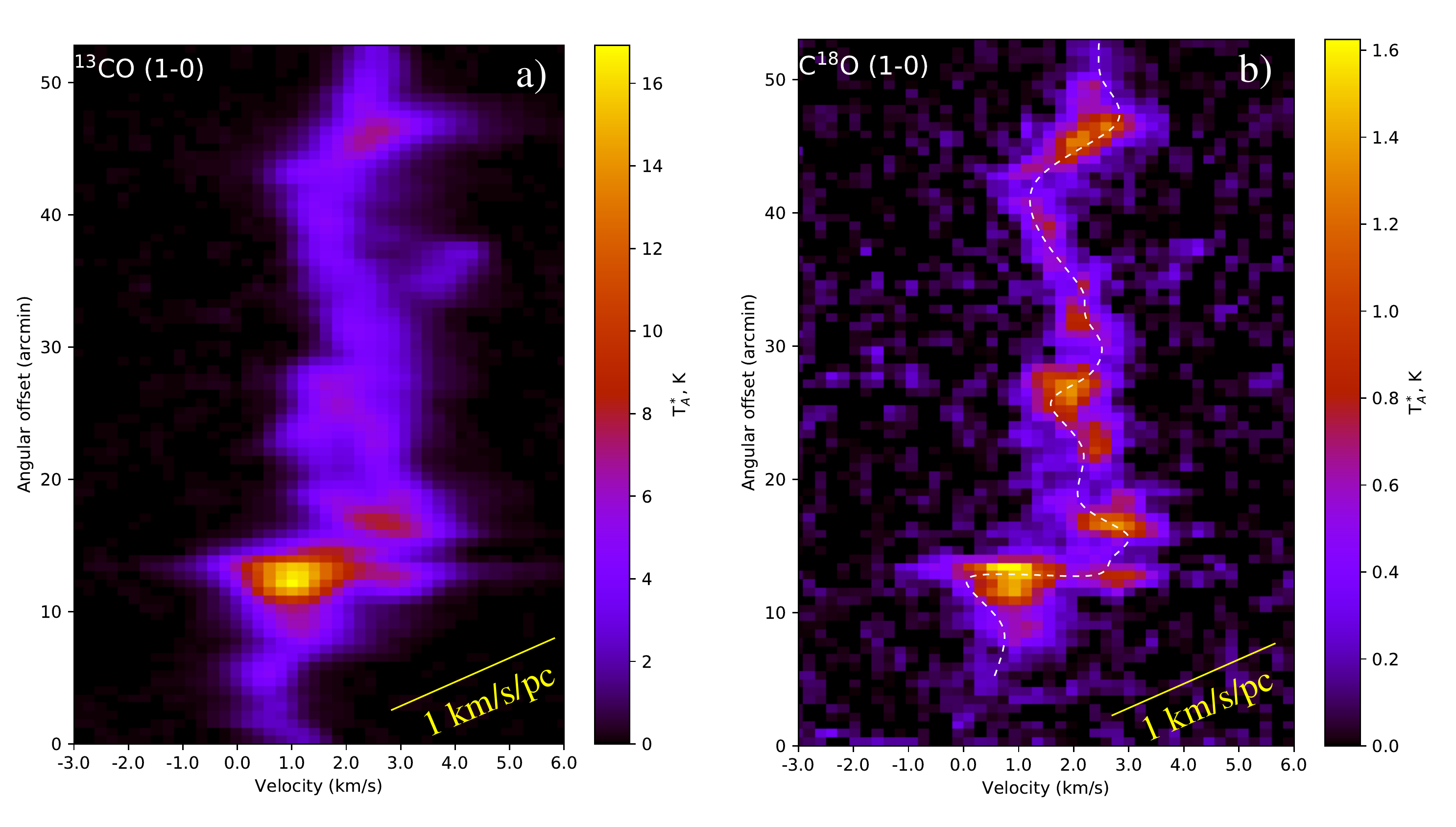}
\caption{a) The $^{13}$CO(1-0) position-velocity plot. 
Antenna temperatures are shown in colors, and represent the averaged values for the bins with 
dimensions 40$''$ $\times$ 120$''$ along a solid curve shown in Figure~\ref{fg8}a. 
b) Same as Figure~\ref{fg9}a, but for C$^{18}$O(1--0). 
An arbitrarily chosen dashed curve (in white) shows an oscillatory-like velocity pattern along the filamentary structure. 
A scale bar corresponding to 1 km s$^{-1}$ pc$^{-1}$ is shown in each panel.}
\label{fg9}
\end{figure*}

In order to study the velocity structure of molecular gas in our selected target, the pv maps of $^{13}$CO and C$^{18}$O emission are shown in 
Figures~\ref{fg9}a and~\ref{fg9}b, respectively. These maps are extracted along the major axis of the filament (see Figure~\ref{fg8}a). 
Both the pv maps reveal an oscillatory-like velocity pattern along the filament, which is more prominently seen in 
Figure~\ref{fg9}b (see a dashed curve in Figure~\ref{fg9}b). Interestingly, velocity gradients ($\sim$1 km s$^{-1}$ pc$^{-1}$) are also observed in the direction 
of the both filament ends, where velocity oscillations are quite large compared to other parts. 
Furthermore, the velocity spread (i.e., $-$1 to 4 km s$^{-1}$) is also very high in 
the direction of the S242 H\,{\sc ii} region or IRAS 05490+2658. 
Previously, \citet{snell90} found a molecular CO outflow toward IRAS 05490+2658 (see Figure~11 in their paper).

\begin{figure*}
\epsscale{0.85}
\plotone{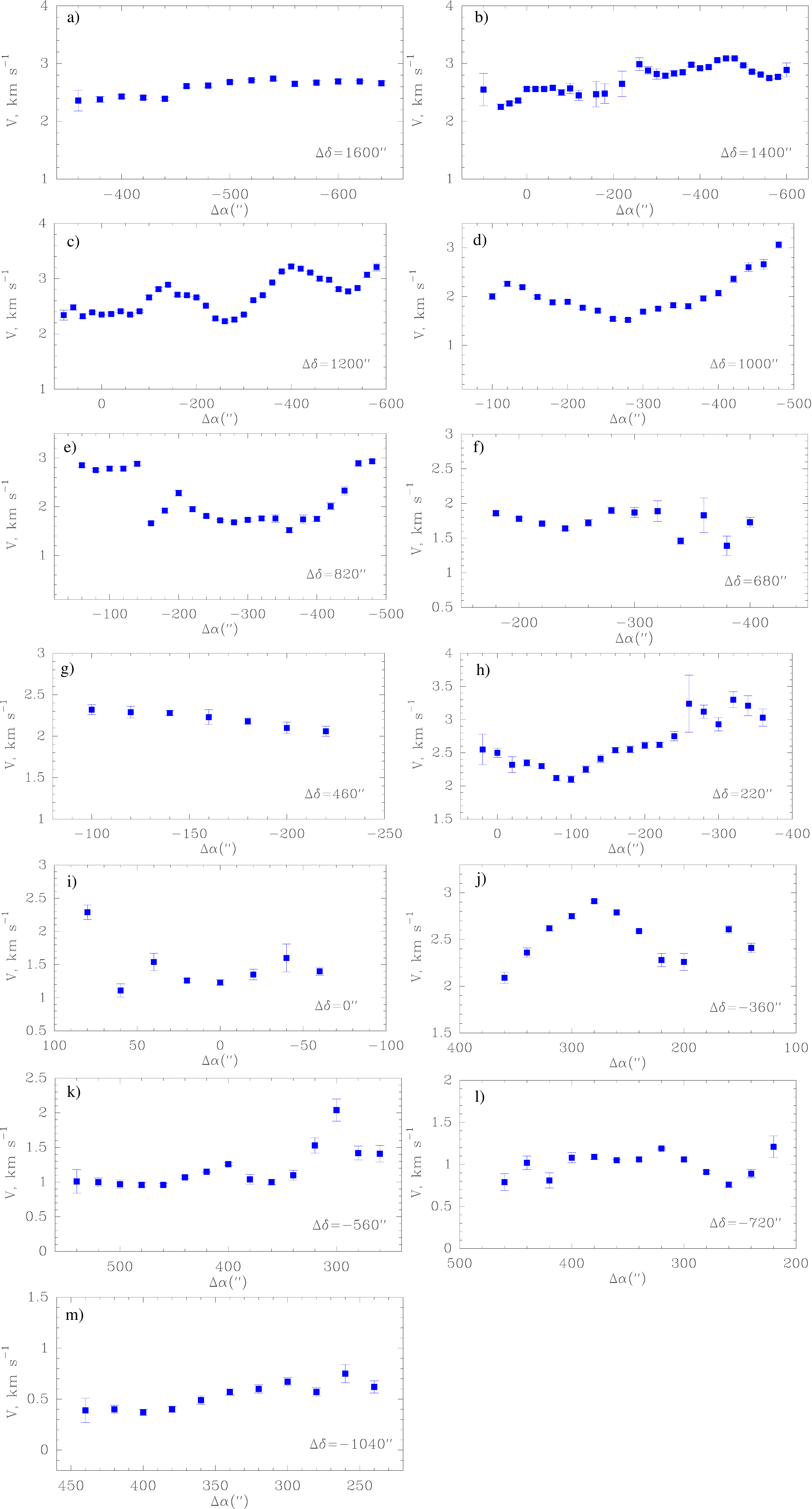}
\caption{Plots of velocity scans perpendicular to the filament for different $\Delta \delta$ values (see Figure~\ref{fg2}). In each panel, the separation of 100$''$ corresponds to 1 pc (at a distance of 2.1 kpc).} 
\label{fg7}
\end{figure*}

\begin{figure*}
\epsscale{0.71}
\plotone{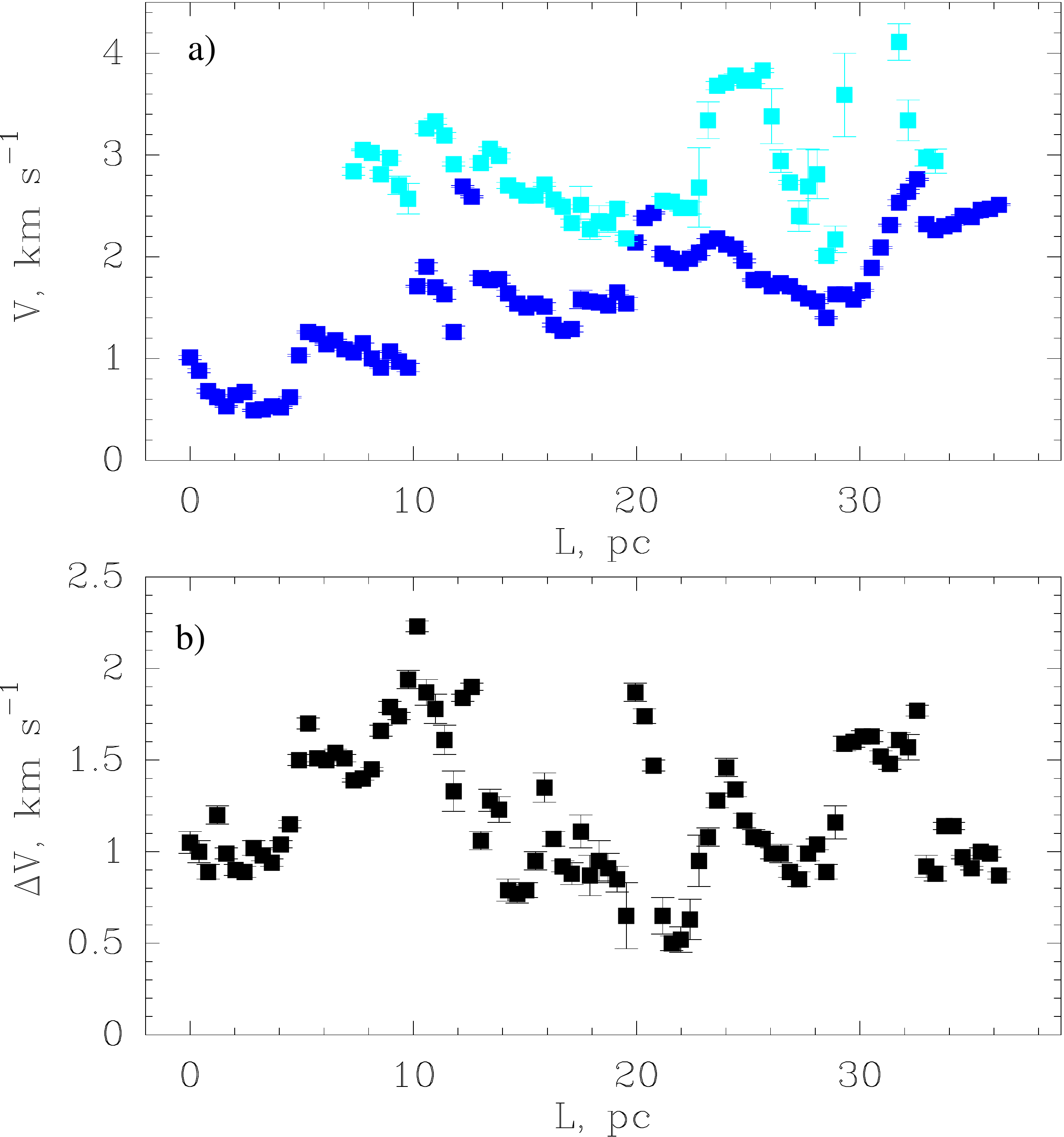}
\epsscale{0.71}
\plotone{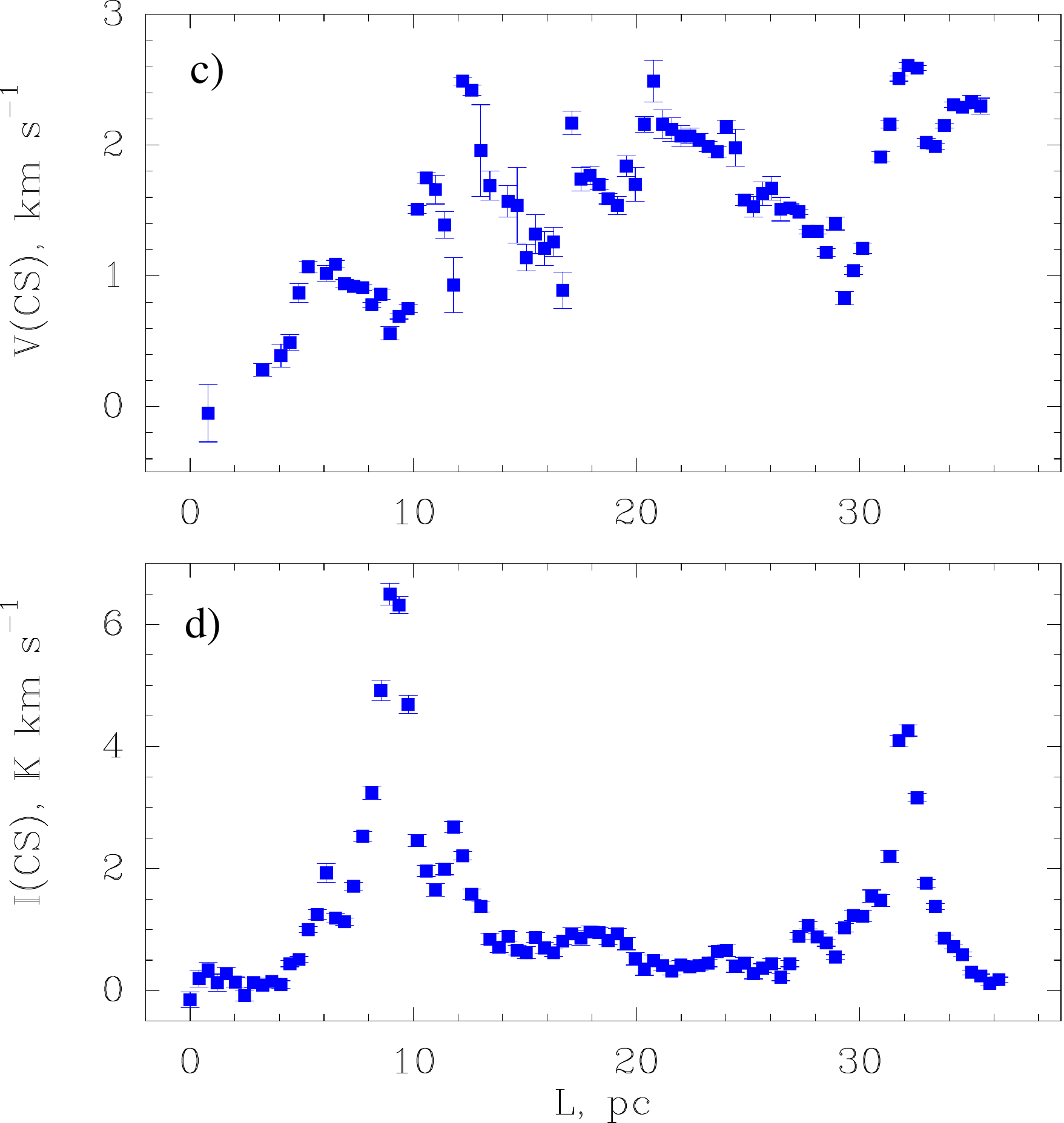}
\caption{a) The variation of $^{13}$CO V$_{lsr}$ along the filament (see a solid curve in Figure~\ref{fg8}a). 
Both the main V$_{lsr}$ (dark blue) and the second V$_{lsr}$ component (light blue) are presented against the length. b) Same as Figure~\ref{fg10}a, but for $^{13}$CO $\Delta$V (i.e. FWHM of the main component). c) The variation of CS(2--1) V$_{lsr}$ along the filament (see a solid curve in Figure~\ref{fg8}a). 
d) The variation of CS(2--1) integrated intensity along the filament (see a solid curve 
in Figure~\ref{fg8}a). In each panel, all the points are obtained from the averaging of 9 
spectra over the same bins. In the panels ``c" and ``d", each data point is obtained from the CS(2--1) spectra, which are averaged over the 
bin with dimensions of 60$''$ $\times$ 60$''$.}
\label{fg10}
\end{figure*}

In Figure~\ref{fg7}, we present plots of velocity scans nearly perpendicular to the filament for different $\Delta \delta$ values 
(see also Figure~\ref{fg2}). In several panels in Figure~\ref{fg7}, one can find signatures of the oscillatory-like velocity pattern, and 
the velocity is more or less constant in some panels. 
Figures~\ref{fg10}a and~\ref{fg10}b display a variation of $^{13}$CO V$_{lsr}$ and $\Delta$V against 
the major axis of the filament, respectively. In order to produce these figures, we analyzed the spectra, where the $^{13}$CO 
integrated intensities are higher than 3$\sigma$. 
Based on a visual inspection of the total spectral map, emission region was divided into subzones. 
For each subzone, the spectra were fitted with one or two Gaussian functions. 
An initial guess of line parameters was taken from averaged spectra over the subzones. 
Figures~\ref{fg10}c and~\ref{fg10}d show a variation of CS(2--1) V$_{lsr}$ and CS(2--1) integrated intensity against 
the major axis of the filament, respectively. In Figure~\ref{fg10}, all the points are obtained from the averaging of 9 spectra over the same bins. 

\begin{figure*}
\epsscale{0.54}
\plotone{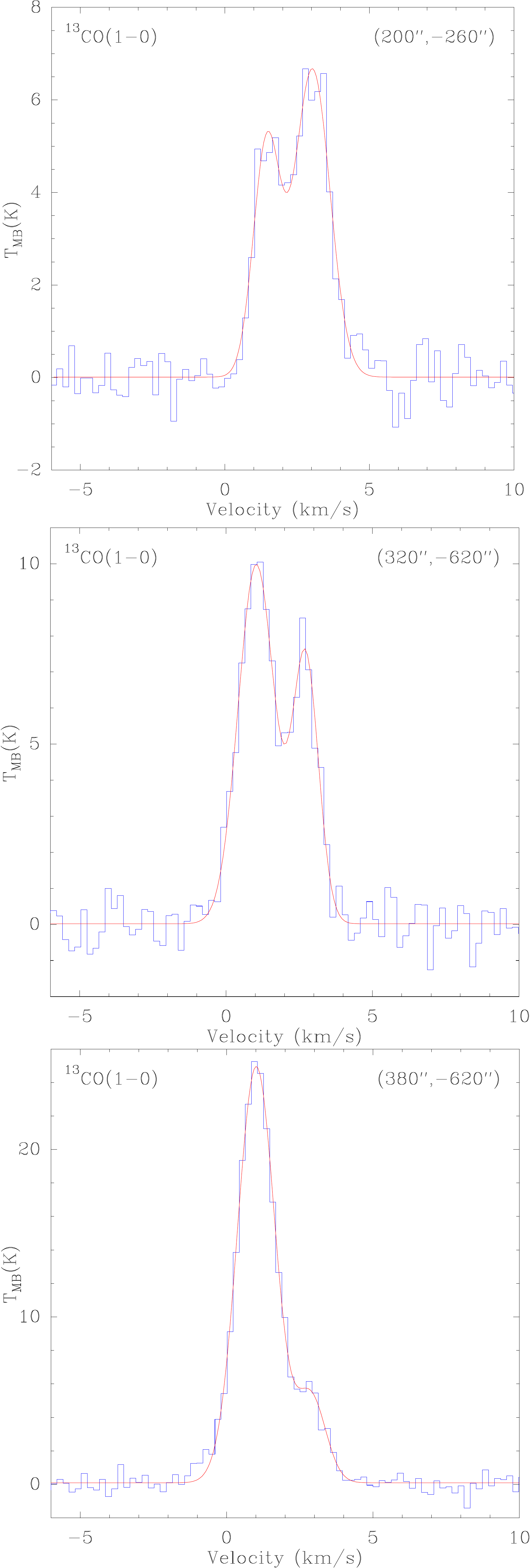}
\caption{Two-component $^{13}$CO spectra at different positions (see also Figure~\ref{fg2}). The position offsets are marked at top right in all the panels. In each panel, the observed spectrum is overlaid with the Gaussian fit.}
\label{xtfg7}
\end{figure*}

We have fitted the Gaussian profile(s) to the observed individual spectrum to model line shape, which enables us to infer the value of 
radial velocity ($^{13}$CO V$_{lsr}$) and linewidth ($^{13}$CO $\Delta$V). 
Figure~\ref{xtfg7} shows three $^{13}$CO spectra at different positions, which are also superimposed with the Gaussian profiles. 
The $^{13}$CO line profiles in the main part of the filament (i.e., $\sim-$700$''$ $<$ $\Delta\delta$ $<$ $\sim$ 1000$''$) 
are often double-peaked. The C$^{18}$O(1-0) profiles also show double-peaked behaviour 
as seen in the $^{13}$CO line profiles, implying superposition of at least two closely located velocity components instead of systematic motions. In many cases, two components are clearly 
separated by 1 km s$^{-1}$, where two Gaussian functions were fitted to the spectra (see Figure~\ref{xtfg7}). In the remaining cases, a single Gaussian fitting was performed. 
The component with lower velocity corresponds to the main velocity component of the filament. 

In Figure~\ref{fg10}a, we have plotted main velocity component with dark blue squares, while the second velocity component is drawn with 
light blue squares. The FWHM of the main component ($\Delta$V) against the major axis of the filament is shown in 
Figure~\ref{fg10}b. An oscillatory-like pattern with a period of $\sim$6--10 pc is seen in both the figures. In Figure~\ref{fg10}d, there are two peaks seen prominently, which correspond to the cores ``1" and ``3", respectively. It seems that an oscillatory pattern in CS(2--1) V$_{lsr}$ could be connected with the fragmentation at the ends of the filament, while in the case of the central part, the CS(2--1) data of higher quality is needed to make definite conclusions related to the link of velocity oscillations with fragmentation (see Figures~\ref{fg10}c and~\ref{fg10}d). 

The implication of all these results are discussed in Section~\ref{sec:disc}.
\subsubsection{Structure function in velocity}
\label{xxssec:data3}
In Section~\ref{ssec:data3x}, we find that the observed linewidths are much higher than the 
thermal ones, implying that the non-thermal (such as turbulent motions) could be responsible for the line broadening in the filament. 
With the help of our $^{13}$CO line data, it is possible to employ statistical methods for analyzing the properties of 
turbulence in our selected target.
The studies of turbulence in molecular clouds usually use the dependence between molecular linewidths and cloud sizes. 
In this connection, in the literature, we find the Larson's one-dimensional velocity dispersion-size relationship with 
$\delta V$ = 0.63 $\times$ L$^{0.38}$, which was obtained for a range of cloud sizes of $\sim$0.1--100 pc \citep{larson81}. 
Using more homogeneous data sets, \citet{solomon87} corrected 
the slope of the Larson's dependence to be 0.5.
Later, velocity structure functions of individual sample clouds derived by \citet{heyer04} were appeared to be nearly 
identical to Larson's $\delta V$--L dependence, indicating that it has its origin in turbulence properties.
For the Musca cloud, \citet{hacar16} derived velocity structure function using the $^{13}$CO data, and found 
significant departures from the Larson's law. They explained this deviation by the existence of sonic-like structures in the cloud decoupled 
from the supersonic turbulent regime.

\begin{figure*}
\epsscale{1}
\plotone{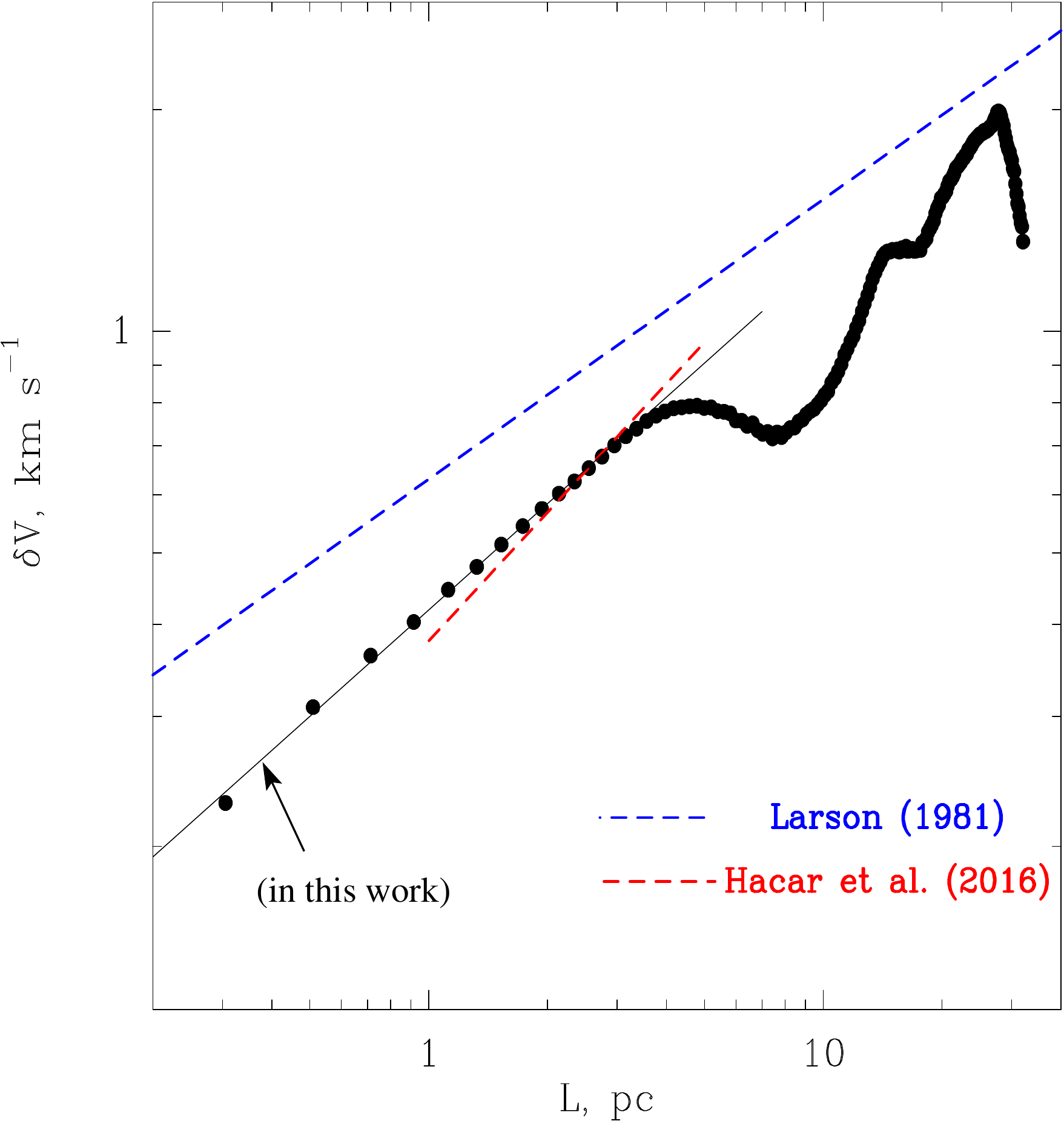}
\caption{Structure function in velocity $\delta V$ as a function of length (i.e. lag) 
derived from the $^{13}$CO line data. The structure function is derived using the total data set.
The Larson's velocity dispersion-size relationship (i.e. $\delta V$ = 0.63 $\times$ L$^{0.38}$) is also marked by a broken blue line. 
The velocity dispersion-size relationship of the Musca cloud is also shown by a broken red line \citep[e.g., $\delta V$ = 0.38 $\times$ L$^{0.58}$;][]{hacar16}. In the filament, a linear relation is found between log($\delta V$) and log(L) for L $\la$ 3 pc, 
where $\delta V$ = 0.42 $\times$ L$^{0.48}$ (see a solid black line in the figure and also text for more details). The errors in the intersection coefficient and the slope are $\sim$0.003 and $\sim$0.01, respectively.} 
\label{fg11}
\end{figure*}

Following the analysis presented in \citet{hacar16}, we calculated the reframed/second-order structure function ($S_2(L)$) using the $^{13}$CO main velocity component (see blue points in Figure~\ref{fg10}a), and the square root of the second-order structure function is given by:
\begin{equation}
S_2(L)^{1/2} = \delta V=  \left<|V(r)-V(r+L)|^2\right>^{1/2}
\label{xxshh1}
\end{equation}
In the calculations, the $^{13}$CO integrated intensities (I($^{13}$CO(1-0))) $\geq$ 2 K km s$^{-1}$ are adopted, and 
a total number of the data points are found to be 2346. 
The total range of angular distances ($\sim$20$''$ -- 3170$''$) between two positions is 
divided into subranges (or lags) of 20$''$ width ($\sim$0.2 pc in linear scale).
For each lag, the value of structure function in velocity (i.e., $\delta V$) is calculated. 
The error of the structure function for a given lag was calculated by individual errors of line velocities obtained from the Gaussian fits and by the method of error propagation. 
It is found to be $\la 0.001$ km s$^{-1}$ for the most cases.
The result concerning the structure function in velocity calculated for the total data set
is shown in Figure~\ref{fg11}, where angular distances are converted into linear ones (L). 
For L $\la$ 3 pc, we find a power-law behavior (i.e., $\delta V$ = 0.42 $\times$ L$^{0.48}$; see a solid black line in Figure~\ref{fg11}). 
The uncertainties of the regression line parameters are small, and the correlation coefficient is close to unity. 
This observed dependence lies lower than the Larson's dependence (see a broken line (in blue) in Figure~\ref{fg11}). 
However, the observed slope in this study is close to the one found by \citet{solomon87}. 
Furthermore, the behavior of $\delta V$ against length (lag) (or the structure function) is very similar to those found for the Musca cloud in a range of L $\sim$1--3 pc \citep[see Figure~5 in][]{hacar16} (see a broken line (in red) in Figure~\ref{fg11}). For larger distances (L $>$ 3 pc), the structure function rises, and has an oscillatory behavior. 
We have calculated the structure function excluding the regions of active star formation (i.e., the northern and southern filament ends), however we do not find any difference from the result obtained for the total data set. 
We have also made a comparison of our outcomes with the published results of the Musca cloud \citep{hacar16} and, have found 
no prominent deviation from the single power-law dependence at lower lags.

The results related to the structure function apparently depend on the inclination effect. 
If the filament has non-zero inclination then the projection of the linear distance between two points, which is collinear to the projection of the filament's axis, will be increased by a factor of 1/cos(i), where ``i" is an inclination angle. 
This will cause stretching of the structure function compared to the case of a zero inclination. 
The stretching is expected to be non-uniform. For the lags much larger than the filament's width, the projections of 
the distances between two points will be nearly collinear to the projection of the filament's axis. 
For small lags, an averaging will go over projections with different angles with respect to the projection of 
the filament's axis, and the stretching will be lower.
However, even in this case, the stretching will take place, and a slope of the regression line calculated for the range of small lags 
will be lower than in the case of a zero inclination. To obtain the precise estimates, a detailed modelling is needed.

We have also calculated the structure functions for different sub-regions as outlined in Figure~\ref{fg8}b. 
For small lags ($\la$ 2--3 pc), the structure functions of the sub-regions depend on L, illustrating the power-law behaviors. 
However, the intersection coefficients and slopes of the regression lines are not the same, and vary in the ranges $\sim$0.3--0.5 and $\sim$0.4--0.6, respectively, depending on the sub-region. 
In general, this could be due to different inclinations as stated above. 
It also includes the contributions of non-turbulent systematic motions (such as, infall and rotation) to velocity structure of sub-regions. 
Furthermore, all of the dependencies lie lower than the Larson's one. 
However, for higher lags, the power-law dependence breaks, and the oscillating mode dominates.

If we take the data for the entire filament then the variations in the structure function parameters of the sub-regions 
are averaged out for compensating individual differences except the global velocity gradient along the filament. 
In order to examine the influence of global linear gradient to our results, 
we derived the structure function for the data along the solid curve (see Figure~\ref{fg10}a), 
and compared this result against the structure function computed for the data where linear velocity gradient was subtracted. 
There are no significant differences found in the parameters of regression lines for lags $\la$ 3 pc.
For higher lags, any dependence of the structure function on L disappears.

The observed power-law dependence could indicate some general behaviour of the velocity field of the gas in the filament. 
In particular, it could be a result of the supersonic vortices produced by gas flows along the filament, which dominate at lower lags.
At higher lags ($L\ga 2-3$~pc), contributions from large-scale 
and ordered velocity gradients and/or filament fragmentation seem to be dominated. Hence, we could conclude that the value of $\sim$3 pc could be an upper scale of the turbulent vortices in the S242 filament.
To confirm these conclusions and to further study the properties of turbulence, it is important to derive the structure 
functions of other molecular clouds and filaments, which will also allow for a comparison.
\section{Discussion}
\label{sec:disc}
In order to understand the physics of 
filament fragmentation and connection of filaments to star formation mechanisms, {\it Herschel} sub-mm images in conjunction 
with observations of molecular line data sets have been employed. Such approach is a powerful tool to assess the ongoing physical processes in the selected target site. 
To examine the dynamical state of a filament, its observed line mass is compared with the critical line mass, 
enabling one to infer its stability \citep[e.g.,][]{andre10,kainulainen16,williams18}. 
It has been highlighted in the literature that filaments can collapse longitudinally along their main axis (i.e. global collapse). 
Two basic types of processes concerning the freefall collapse/global collapse of a uniform-density filament are reported, which are the homologous collapse for the filament with aspect ratio of A $\leq$ 5 
and the end-dominated collapse for the filament with A $\geq$ 5 \citep[e.g.,][]{pon12}. For filamentary clouds, the aspect ratio 
is defined as A = {\it l} / R, where {\it l} is the filament's half-length and R is its radius \citep[see][for more details]{toala12}.  
It has also been pointed out that the elongated filaments with large aspect ratios are more prone to end-dominated collapse \citep[e.g.,][]{pon12}. As mentioned in Introduction, observational examples of end-clumps in isolated filaments are limited \citep[e.g.,][]{zernickel13,beuther15,kainulainen16,hacar16}.

\begin{figure*}
\epsscale{0.78}
\plotone{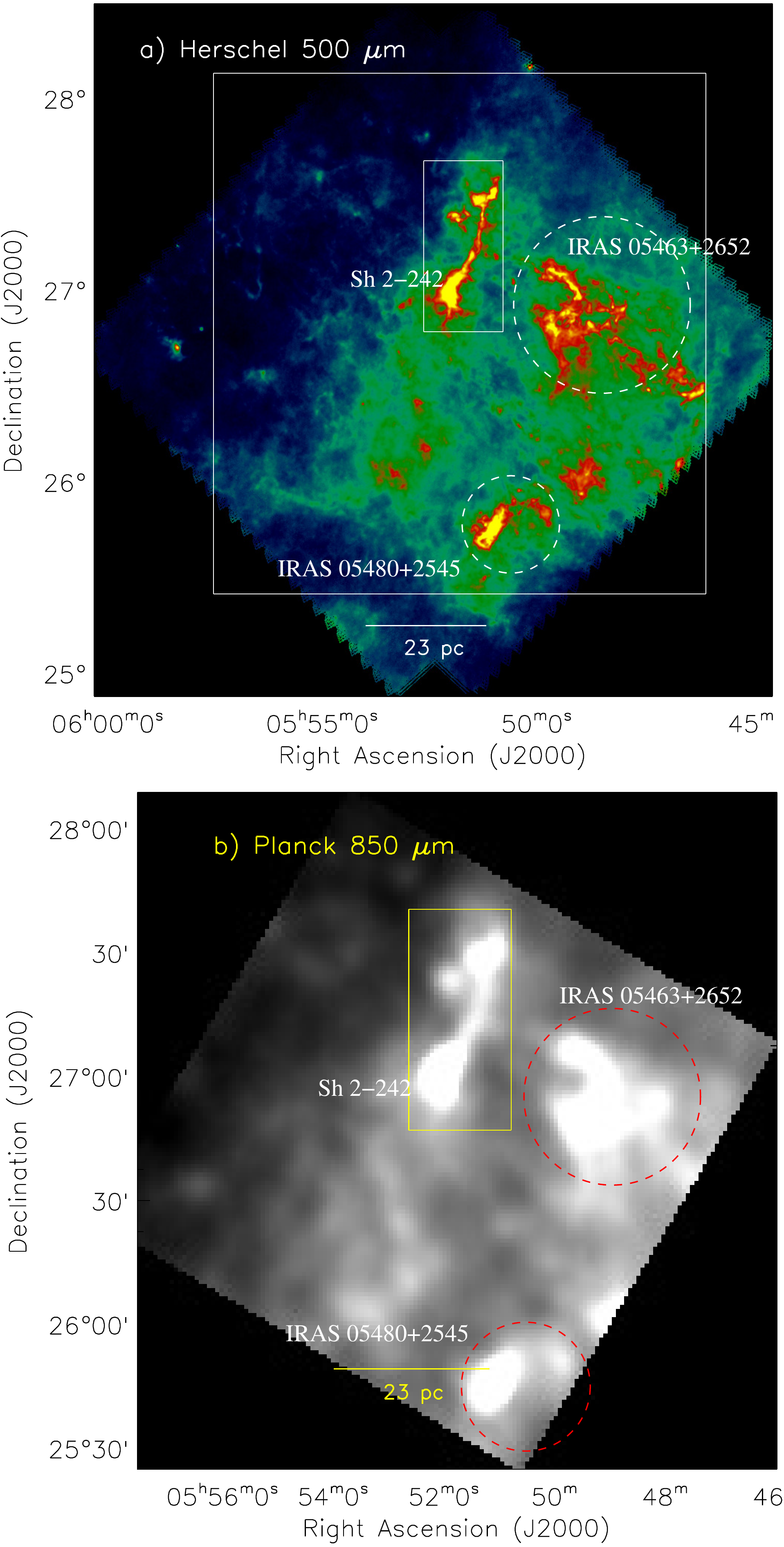}
\caption{Large-scale image of a field hosting the site S242. 
a) False color {\it Herschel} image at 500 $\mu$m. 
The square box highlights an area of 2$\degr$.8 $\times$ 2$\degr$.8 (or 102.6 pc $\times$ 102.6 pc). 
b) Grey-scale color {\it Planck} image at 850 $\mu$m (see a square box in Figure~\ref{xxfg11}a). 
In each panel, a rectangle box indicates the area shown in Figure~\ref{fg1}, and also highlights 
the location of the S242 filament. 
In both the panels, broken circles show the star-forming sites IRAS 05480+2545 \citep{dewangan17b} and IRAS 05463+2652 \citep{dewangan17d}.} 
\label{xxfg11}
\end{figure*}

In Paper~I, the selected target S242 filament has been proposed as the end-dominated collapse candidate 
using the continuum observational data sets. Using the sub-mm images at {\it Herschel} 500 $\mu$m (resolution $\sim$37$''$) and {\it Planck} 850 $\mu$m (resolution $\sim$300$''$), 
Figures~\ref{xxfg11}a and \ref{xxfg11}b display a large-scale area ($\sim$102 pc $\times$ 102 pc) containing the site S242 \citep[$^{13}$CO V$_{lsr}$ $\sim$0.7 km s$^{-1}$;][]{kawamura98}, respectively.  
In these figures, two star-forming sites, IRAS 05480+2545 \citep[$^{13}$CO V$_{lsr}$ $\sim$$-$9.1 km s$^{-1}$;][]{kawamura98,dewangan17b} and 
IRAS 05463+2652 \citep[$^{13}$CO V$_{lsr}$ $\sim$$-$10.6 km s$^{-1}$;][]{kawamura98,dewangan17d}, are highlighted by broken circles. 
Considering the values of $^{13}$CO V$_{lsr}$, these two IRAS sites do not appear to be physically connected with the site S242 
\citep[see S242 region around $l$ = 182$\degr$.40; $b$ = 0$\degr$.27 in Figure 9l in][]{kawamura98}. 
In the dust continuum images, the S242 filamentary structure appears as an isolated molecular cloud (see a rectangle box in Figures~\ref{xxfg11}a and \ref{xxfg11}b), 
and its both ends are not linked with the other nearby regions. 

With the aid of new OSO-20m molecular (CO and CS) maps, several supersonic molecular sub-regions are identified in the filament (see Section~\ref{ssec:data3x}). 
In Section~\ref{ssec:data3x}, with the examination of the dynamical states, the selected sub-regions (except sub-region ``2") in the filament are found to be supercritical, and could not be supported by a combination of thermal and turbulent motions. 
Furthermore, dense molecular cores are observed mainly toward both the filament ends, and are close to virial equilibrium 
(see Section~\ref{ssec:data3x} and also Table~\ref{tab2}). Additionally, both the ends of the filament contain molecular cores with almost similar masses (see Table~\ref{tab2}). In Paper~I, the dynamical or expansion age of the S242 H\,{\sc ii} region was obtained 
to be $\sim$0.5--1.8 Myr (for n$_{0}$ = 10$^{3}$--10$^{4}$ cm$^{-3}$). In general, the average lifetime of YSOs (Class~I and Class~II) can be considered to 
be $\sim$0.44--2 Myr \citep[e.g.,][]{evans09}. 
\citet{clarke15} estimated a collapse timescale (t$_{col}$) for the filamentary structures, which is defined as t$_{col}$ = (0.49 + 0.26A)/(G$\rho)^{1/2}$, where A is the initial aspect ratio of the filament, G is the gravitational constant, and $\rho$ is the volume density of the filament. In the present case, using the molecular line data, the radius, aspect ratio, and density of the 
filament can be estimated. The observed aspect ratio of the S242 filament is computed to be A $\sim$ (30 pc/2)/(1.5 pc/2) $\sim$20. 
Using a mean density of n = 10$^{4}$ cm$^{-3}$ (or $\rho$ = 2.33 $\times$ 1.67 $\times$ 10$^{-24}$ $\times$ 10$^{4}$ gm cm$^{-3}$; see Table~\ref{tab2}), we obtain the collapse timescale of t$_{col}$ $\sim$3.5 Myr, which is much older than the formation of massive stars and clusters of YSOs in the filament. This argument is still acceptable even if we assume about 10--20\% error in the above estimates.

The pv plots of molecular lines reveal velocity gradients ($\sim$1 km s$^{-1}$ pc$^{-1}$) in the direction of both the filament ends. 
One of the possibilities to explain the observed velocity gradients is star formation activities, while the other 
explanation of the velocity gradients in the filament is due to the acceleration of the gas.
Earlier in the literature, based on the observed velocity gradients along filaments, gas flows along filaments were proposed in both nearby low-mass star-forming clouds \citep{hacar11,kirk13} and 
massive clouds \citep{schneider10,peretto14,tackenberg14}. 
Furthermore, the oscillatory pattern in velocity is also observed in the velocity space of molecular lines (see also Figure~\ref{fg10}). 
\citet{hacar11} observationally studied the filament L1571, and carried our a modelling of velocity oscillations as
sinusoidal perturbations. They suggested that the observed velocity oscillations along the filament may indicate the filament 
fragmentation process via accretion along filament. 
In Section~\ref{xxssec:data3}, the structure function in velocity of the filament is derived using the $^{13}$CO line data. This analysis indicates that the velocity field in the filament could 
consist of both turbulent random motions, which dominate on lower- and large-scales, and ordered velocity gradients for $L\ga 2-3$~pc.
It also suggests the filament fragmentation, which is probably caused by gravitational contraction/accretion 
as well as by other feedback effects on the local scale \citep[e.g., ionizing radiation, stellar winds, 
and star formation activities;][]{arzoumanian13}. 
The differences between power-law dependencies found for various sub-regions of the filament
and from the Larson's law could be in general connected with local properties 
of turbulence in S242. To confirm this conclusion, more data for other objects are needed.
Hence, in the case of our selected target, the oscillatory pattern in velocity is suggestive of fragmentation 
process through accretion along the S242 filament. 

Taken together, we find that the observed results of the S242 filament 
are in agreement with the end-dominated collapse scenario.
\section{Summary and Conclusions}
\label{sec:conc}
To understand the physical processes in the S242 filament, we have examined its kinematic structure using CO ($^{13}$CO(1--0) and C$^{18}$O(1--0)) and CS(2--1) line emission data, which were observed with the OSO-20m telescope. 
The major results of the present work are as follows:\\

$\bullet$ The molecular cloud associated with the S242 filament is studied in a velocity range of [$-$1, 5] km s$^{-1}$, and its elongated nature (having length $\sim$30 pc) is also traced in the molecular maps. One of the filament ends contains the S242 H\,{\sc ii} region, 
while the other end hosts several {\it Planck} cold clumps.\\ 
$\bullet$ Several sub-regions are selected in the filament. 
The values of Mach number ($\sigma_{NT}/c_s$) derived using the molecular line data for the sub-regions are 
computed to be about 2.7--4.0, suggesting these sub-regions are supersonic.\\ 
$\bullet$ The observed $M_{\rm line,obs}$ of the sub-regions (except central part) in 
the filament are larger than their $M_{\rm line,vir}$, indicating that these sub-regions (except central part) are supercritical.
Hence, the supercritical sub-regions in the filament may not be supported by a combination of thermal and turbulent motions.\\ 
$\bullet$ Dense cores are detected in the CS and C$^{18}$O maps, and are seen mainly toward both the filament ends.
These cores are also close to virial equilibrium.\\ 
$\bullet$ Mean volume densities of the dense cores at the southern end of the filament are several times
higher than those at the northern end.
Both the filament ends contain molecular cores with almost similar masses. 
The filament's collapse time is estimated to be $\sim$3.5 Myr.\\
$\bullet$ Noticeable YSOs are found toward the entire filament, but they are more concentrated toward its ends.\\
$\bullet$ In the direction of the filament, an oscillatory pattern in velocity is evident, which is indicative of fragmentation. 
The pv plots of $^{13}$CO and C$^{18}$O reveal velocity gradients ($\sim$1 km s$^{-1}$ pc$^{-1}$) toward both the filament ends.\\
$\bullet$ Using the $^{13}$CO line, the structure function in velocity of the filament is computed, which is 
similar as found in the Musca cloud for lags $\sim$1--3~pc. 
In the filament, the structure function rises with distance, and shows a power-law behavior 
(i.e. $\delta V$ = 0.42 $\times$ L$^{0.48}$) at L$\la$~3~pc and an oscillatory pattern at higher L.
This dependence lies lower than the Larson's velocity dispersion-size relationship.
The oscillatory pattern in velocity at higher lags is suggestive of large-scale and ordered velocity gradients and
fragmentation process via accretion along the S242 filament.\\

All the observational results along with their uncertainties put together, the S242 filament is found to be a reliable candidate of the end-dominated collapse.
\acknowledgments
We thank the anonymous reviewer for a critical reading of the manuscript and several useful comments and 
suggestions, which greatly improved the scientific contents of the paper.  
The research work at Physical Research Laboratory is funded by the 
Department of Space, Government of India. 
The OSO-20m observations and data processing were done under support of the Russian Foundation for Basic Research (RFBR) (projects 17-52-45020 and 18-02-0060). The OSO-20m data analysis were done under support of the Russian Science Foundation (project 17-12-01256).
%
\begin{deluxetable}{cccccccc}
\tablewidth{0pt} 
\tabletypesize{\scriptsize} 
\tablecaption{Physical parameters of different molecular sub-regions as highlighted in Figure~\ref{fg8}b. 
Column~1 gives the IDs assigned to the sub-regions. Table also lists 
length, C$^{18}$O mass (M$_{subreg}$), line mass ($M_{\rm line,obs}$), FWHM (C$^{18}$O $\Delta V$), non-thermal velocity component ($\sigma_{NT}$), and critical line mass ($M_{\rm line,vir}$). These sub-regions are distributed toward the filament. \label{tab1}} 
\tablehead{ \colhead{ID} & \colhead{Length}& \colhead{M$_{subreg}$} & \colhead{$M_{\rm line,obs}$}& \colhead{$\Delta V$ (C$^{18}$O)}& \colhead{$\sigma_{NT}$}& \colhead{$M_{\rm line,vir}$ (at T = 10 K)}\\
\colhead{} & \colhead{(pc)} &\colhead{($M_\odot$)} & \colhead{($M_\odot$/pc)}& \colhead{(km s$^{-1}$)}& \colhead{(km s$^{-1}$)}& \colhead{($M_\odot$/pc)}}
\startdata 
          1  &   9        &	  2353  & 261 &1.48 & 0.63 & 192 \\	  
          2  &   16	  &	  3351  & 209 &2.01 & 0.85 & 338 \\
          3  &   4.5	  &	  957   & 213 &1.22 & 0.52 & 135 \\
          4  &    7	   &	  1952  & 279 &1.29 & 0.55 & 149 \\
          5  &    5	   &	  1045 & 209 &1.30 & 0.55 & 153 \\
\end{deluxetable}
\begin{deluxetable}{cccccccc}
\tablewidth{0pt} 
\tabletypesize{\scriptsize} 
\tablecaption{Five dense cores observed in the CS intensity map (see Figure~\ref{fg8}c). 
Column~1 gives the IDs assigned to cores. Table also lists 
central positions, CS diameter (D$_{c}$), mass derived from C$^{18}$O (M$_{c}$), FWHM (CS $\Delta V$), M$_{vir}$, and mean number density
calculated from M$_c$ and D$_c$. \label{tab2}} 
\tablehead{ \colhead{ID} & \colhead{{\it RA}} & \colhead{{\it Dec}} & \colhead{D$_{c}$}& \colhead{M$_{c}$} & \colhead{$\Delta V$}& \colhead{M$_{vir}$}
& \colhead{$\bar n$}\\
\colhead{} &  \colhead{[2000]} & \colhead{[2000]} & \colhead{(pc)} &\colhead{($M_\odot$)} &\colhead{(km s$^{-1}$)} &\colhead{($M_\odot$)}&\colhead{(cm$^{-3}$)}}
\startdata 
          1  &   5:52:09.3     &  +27:00:23.3        &    0.76     &	  295  & 2.27  & 409 &22500 \\	  
          2  &   5:52:01.6     &  +27:03:29.1        &    0.92	   &	  272  & 2.04  & 404 &11600 \\
          3  &   5:51:32.1     &  +27:28:55.5        &    1.33	   &	  345  & 1.53  & 325 &4900 \\
          4  &   5:51:11.6     &  +27:31:47.9        &    1.61	   &	  250  & 1.32  & 295 &2010 \\
          5  &   5:51:59.6     &  +27:23:51.6        &    1.39	   &	  271  & 1.18  & 204 &3400 \\
\end{deluxetable}
\end{document}